\documentstyle[11pt,aaspp4]{article}

\def\sn{SN~1987A}
\def\iue{{\it IUE\/}}
\def\hst{{\it HST\/}}

\def\Ha{${\rm H}\alpha$}
\def\Hb{${\rm H}\beta$}
\def\Hg{${\rm H}\gamma$}

\def\EE#1{\times 10^{\small#1}}

\def\cm3{\rm ~cm^{\small -3}}
\def\kms{\rm ~km~s^{\small -1}}

\def\ergcms{\rm\,ergs~cm^{\small -2}~s^{\small -1}}

\def\wl{$\lambda$}
\def\wll{$\lambda\lambda$}
\def\hi{{\ion{H}{1}}}

\def\hei{{\ion{He}{1}}}

\def\cii{{\ion{C}{2}}}
\def\ciii{{\ion{C}{3}}}
\def\civ{{\ion{C}{4}}}
\def\nii{{\ion{N}{2}}}
\def\niii{{\ion{N}{3}}}
\def\niv{{\ion{N}{4}}}
\def\nv{{\ion{N}{5}}}
\def\nvi{{\ion{N}{6}}}
\def\oi{{\ion{O}{1}}}
\def\oii{{\ion{O}{2}}}
\def\oiii{{\ion{O}{3}}}
\def\oiv{{\ion{O}{4}}}

\def\ovii{{\ion{O}{7}}}
\def\sii{{\ion{S}{2}}}

\slugcomment{To appear in 1 December 2000 ApJ (Part 1)}
\lefthead{Maran et al.}
\righthead{STIS Spectroscopy of SN 1987A Rings}

\begin{document}

\title{Physical Conditions in Circumstellar Gas surrounding SN 1987A 12 Years After 
Outburst\altaffilmark{1}}

\author{
Stephen P. Maran\altaffilmark{2}, 
George Sonneborn\altaffilmark{3}, 
Chun S. J. Pun\altaffilmark{3,4}, 
Peter Lundqvist\altaffilmark{5},
Rosina C. Iping\altaffilmark{3,6}, 
Theodore R. Gull\altaffilmark{3}, 
}
\affil{}

\altaffiltext{1}{Based on observations with the NASA/ESA {\it Hubble
Space Telescope}, obtained at the Space Telescope Science Institute,
which is operated by the Association of Universities for Research in
Astronomy Inc., under NASA Contract NAS5-26555.}
\altaffiltext{2}{Space Sciences Directorate, Code 600, NASA Goddard Space Flight Center, 
Greenbelt, MD 20771; stephen.p.maran@gsfc.nasa.gov}
\altaffiltext{3}{Laboratory for Astronomy and Solar Physics, Code 681,
NASA Goddard Space Flight Center, Greenbelt, MD 20771; 
pun@congee.gsfc.nasa.gov,
george.sonneborn@gsfc.nasa.gov,
gull@sea.gsfc.nasa.gov, 
rosina@taotaomona.gsfc.nasa.gov}
\altaffiltext{4}{National Optical Astronomical Observatories, 
P.O.Box 26732, Tucson, AZ 85726}
\altaffiltext{5}{Stockholm Observatory, SE-133 36, Saltsj\"obaden, Sweden;
~peter@astro.su.se}
\altaffiltext{6}{Department of Physics, Catholic University of America, 
Washington, DC 20064}

\begin{abstract}

Two-dimensional spectra of Supernova 1987A were obtained on 
1998 November 14-15 (4282 days after outburst) with the Space Telescope 
Imaging Spectrograph (STIS) on board the {\it Hubble Space Telescope\/} (\hst). 
The slit sampled portions of the inner circumstellar ring at the east 
and west ansae as well as small sections of both the northern and southern 
outer rings. The temperature and density at these locations are
estimated by nebular analysis of [\nii], [\oiii], and [\sii] emission line 
ratios, and with time-dependent photoionization/recombination models. 
The results from these two methods are mutually consistent. 
The electron density in the inner ring is $\sim 4000 \cm3$ for \sii,
with progressively lower densities for \nii\ and \oiii. 
The electron temperatures determined from [\nii] and [\oiii] line
ratios are $\sim 11,000$~K and $\sim 22,000$~K, respectively.
These results are consistent with evolutionary trends in the
circumstellar gas from similar measurements at earlier epochs. 
We find that emission lines from the outer rings come from 
gas of lower density ($n_{\rm e} \la 2000 \cm3$) than that which emits the 
same line in the inner ring. 
The N/O ratio appears to be the same in all three rings.
Our results also suggest that the CNO abundances in the northern
outer ring are the same as in the inner ring, contrary to 
earlier results of Panagia et al.\ (1996). 
Physical conditions in the southern outer ring are less certain 
because of poorer signal-to-noise data.
The STIS spectra also reveal a weak \Ha\ emission redshifted by 
$\sim 100 \kms$ at p.a.\ 103\arcdeg\ that coincides with the recently 
discovered new regions that are brightening (\cite{lawr00}).  
This indicates that the shock interaction in the SE section of the inner 
ring commenced over a year before it became apparent in \hst\ images.
\end{abstract}


\section{INTRODUCTION}

Narrow emission lines (FWHM $< 15 \kms$) from nitrogen-enriched 
circumstellar gas around \sn\ were first seen in the UV 
with the \emph{International Ultraviolet Explorer (\iue)} satellite 
in 1987 May (Fransson et al.\ 1989), about 70 days after outburst. 
The circumstellar gas was ionized by the strong burst of extreme UV and 
soft X-ray radiation ($T_{\rm color} \sim 1\EE{6}$ K, \cite{blin00}) 
produced when the shock front of the stellar explosion first broke 
through the surface of the star. 
The circumstellar gas was highly photoionized by the UV flash, 
reaching ioniziation level of at least \nvi\ and \ovii\ (\cite{lund96}).
The subsequent cooling of the gas has produced the ring emission observed 
over the last thirteen years
(except for the shock interaction emission, cf.\ \S\ref{spot3}). 
The UV emission lines peaked in brightness at $\sim$~day 400 when
the light paraboloid swept through the whole inner ring region and these 
lines have been decaying since then (\cite{sonn97}). 
As the emission is dominated by progressively lower density gas, the
flux decay rate decreases with time. 
Optical emission lines (mainly [\oiii]) were first detected on about day~310 
by Wampler \&~Richichi (1989). 
Subsequent ground-based imaging by Crotts, Kunkel, \&~McCarthy (1989)
and Wampler et al.\ (1990) together with 
\emph{Hubble Space Telescope} (\hst) observations with 
the Faint Object Camera (\cite{jako91}) and Wide-Field/Planetary Camera 1
(\cite{plai95}) showed that the brightest 
circumstellar material was distributed in a ring-like structure. 
Following the first \hst\ Servicing Mission, the newly 
stigmatic \hst\ revealed the fainter, more complex structure of the 
circumstellar material surrounding \sn: an equatorial inner ring 
and two outer thin loops were clearly seen (\cite{burr95}). 

The UV emission lines of the inner ring revealed that the \sn\
circumstellar material has enhanced CNO abundances 
(N/C~$\sim 29$ and N/O~$\sim 13$ times solar, \cite{fran89}; \cite{lund96};
\cite{sonn97}).
This has proven to be one of the main observational constraints on 
the evolution of the supernova progenitor, implying that the star was
in a post He-core burning phase at the time of the explosion (\cite{pods92}).
Photoionization/recombination modeling of the early UV line emission 
showed that the peak radiation temperature at the time of shock 
breakout was $\ga 1\EE{6}$ K (\cite{lund96}). 
This temperature is consistent with that predicted by the 
radiation-hydrodynamic modeling of the breakout 
(\cite{ensm92}; \cite{blin00}).
The similar decay times derived for the different ionized states 
of the same element, such as \nv\ and \niii\ (\cite{sonn97}), 
cannot be fitted with a single density model (\cite{lund96}). 
Instead, this result suggests that the inner ring consists of gas
with a wide range of density $n_{\rm e} \sim 7 \EE{3} - 4 \EE{4} \cm3$. 
In this non-equilibrium situation, gas of different density 
dominates the line emission at different times for a particular region
in the ring because of the shorter recombination timescale 
for higher density gas (see \cite{lund96} for a thorough discussion).
In general, the line emission was first dominated by emission from
the highest density gas, and later by successively lower density 
gas within the circumstellar ring. 

The physical conditions of the inner ring have also been 
studied in optical wavelengths both from the ground and with \hst. 
At the time of the earliest optical emission line observations, 
Wampler \&~Richichi (1989) derived a 
mean temperature of 45,000~-- 75,000~K for electron densities 
below $10^5\cm3$ from the [\oiii] \wll4959, 5007/\wl4363 line ratio.
Subsequent analyses (\cite{khan91}; \cite{menz91}; \cite{wang91}) 
of the [\oiii] line ratio up to $\sim$~day~1300 showed
that the temperatures of the emitting O$^{\rm2+}$ ions remained at 
$\sim$~30,000~K, while the [\nii] \wll6548, 6583/\wl5755 line ratio 
indicated a temperature of $\sim$~15,000~K. The electron density 
derived from the [\sii] doublets \wl6716/\wl6731 was $10^{4} \cm3$. 
From \hst/Faint Object Spectrograph observations of the brightest 
location in the inner ring (p.a.\ = 300\arcdeg) taken at day~2876, 
Lundqvist \&~Sonneborn (2000, henceforth LS00) derived the 
[\nii] and [\oiii] temperatures to be $\sim 10,000$~K, and $\sim 27,000$~K, 
respectively, and a [\sii] density of $\sim 9000 \cm3$. 
Combining with the \hst/WFPC2 narrow band F648N ([\nii]\wl6583) and
F502N ([\oiii]\wl5007) observations, the average densities 
around the inner ring of the [\nii]- and [\oiii]-emitting gas 
were determined to be $\ga 4000 \cm3$, and $\sim 2300 \cm3$, respectively.

Compared to the inner ring, the physical conditions of the two
thin outer rings have been much less studied because of their faintness
(surface brightness of the outer rings are $\sim 5 - 15$\% of that
of the inner ring).
Panagia et al.\ (1996) studied UV and optical \hst/FOS spectra of the
position on the northern outer ring that is coincident with the 
position of the \sn\ debris. 
They concluded that while the ioniziation and temperature
derived are similar to that of the inner ring, 
the outer ring gas studied is $\sim 3$ times less CNO enriched 
than the inner ring, with a slightly lower electron 
density $n_{\rm e} \sim 800 \cm3$.
The lower CNO abundances lead Panagia et al.\ to suggest that
the outer ring gas materials were ejected by the \sn\ progenitor 
$\sim 10^4$~years before that of inner ring gas. 
This result contrasts with recent kinematical studies of the
three rings which indicate that all three rings are coeval, 
created $\sim$~20,000 yr before the supernova explosion (\cite{crot00}). 
As shown below in \S\ref{oring}, the line emissions from the outer 
rings are highly time-dependent, and this effect may not have been 
sufficiently taken into account by Panagia et al.\ (1996). 

In this paper we report long slit optical STIS spectroscopy of \hi, [\nii], 
[\oiii], and [\sii] emission from the inner and outer circumstellar 
rings (\S2). Multiple positions on the inner ring and the two outer
rings have been observed and analyzed (\S3). 
Ratios of these transitions form useful plasma diagnostics,
with which we determine $n_{\rm e}$ and $T_{\rm e}$ of the 
emitting gas in the inner (\S\ref{iring}) and outer rings 
(\S\ref{oring}). 
A preliminary version of this analysis was presented by Iping, 
Sonneborn, \&~Pun (1999). 
We constructed time-dependent photoionization/recombination models for 
both the inner and outer rings to check the consistency of the physical
conditions derived from the line ratios. 
Finally, we report in \S\ref{spot3} the detection of a Doppler-shifted 
\Ha\ emission observed in the spectra, which can be attributed to the
onset of interaction between the supernova debris and the ring at 
p.a. = 103\arcdeg. 

\section{OBSERVATIONS} \label{obs}

\sn\ and its rings were observed with the \hst\ Space Telescope Imaging
Spectrograph (STIS) on 1998 Nov 14 and 15 
(4282 days after core collapse).  Medium resolution optical 
spectra taken with gratings G430M and G750M were obtained with a long, 
narrow slit (52\arcsec $\times$ 0\farcs2).
The position of the STIS aperture is shown to scale in 
Figure~\ref{slit-pos}(a). 
The center of the slit is close to the center of the debris and its
orientation of position angle p.a. = 103\arcdeg\ is close
the major axis of the inner ring (p.a. $\sim$ 83\arcdeg, 
Oppenheimer 1999). The spatial resolution in these spectral images 
is $\sim$ 1.4~-- 1.6 pixels, or 0\farcs07~-- 0\farcs08 (FWHM). 

Crotts \& Heathcote (2000; see also Cumming \& Lundqvist 1997) have measured 
the expansion velocities of the inner equatorial ring ($10.5 \pm 0.3 \kms$) 
and the two outer rings ($26 \kms$). The circumstellar rings remain 
spectrally unresolved in the STIS medium resolution grating modes,  
where the spectral resolution is $\sim 50 \kms$. 
Multiple observations ($n=4-5$) centered at dithered positions 
0\farcs5 apart along the slit were made with each grating setting. 
Cosmic rays and hot pixels were removed simultaneously when the 
dithered raw images were combined with the CALSTIS software developed
by the STIS Investigation Definition Team at the Goddard Space
Flight Center\footnote{CALSTIS Reference Guide, 
http://hires.gsfc.nasa.gov/stis/software/doc\_manuals.html}.
The grating settings used and 
the emission lines observed in each setting are listed in Table~\ref{tb-obs}. 

The long slit crossed portions of the inner ring near the
east and west ansae. We refer to these positions in this paper 
as the ``west'' and ``east'' inner ring (WIR and EIR), as labelled
in Figure~\ref{slit-pos}.
Assuming the ring is expanding instead of contracting, 
Crotts \&~Heathcote (1991) have shown that the northern
side of the inner ring is inclined closer towards the observer.
In that scenario, the WIR region is located closer towards
the observer than the EIR region and emissions measured 
from the EIR originate from the inner ring $\sim$~100~days 
{\it before\/} those received simultaneously from the WIR . 

The slit also crosses the northern and southern outer rings each at two 
locations. However, only the emitting gas at the two intersection points 
farther from the supernova has sufficient brightness for our analysis.
We refer to the sections of the outer rings studied in this paper as 
``north outer ring'' (NOR) and ``south outer ring'' (SOR). 
Assuming the same geometry  as the inner ring
(see Figure~3 of \cite{burr95} for an illustration),
then the northern outer ring lies {\it farther\/} from the observer 
than the southern outer ring. And emissions from the NOR region 
originated $\sim$~900~days before those received simultaneously 
from the SOR region. 

\section{EMISSION LINE ANALYSIS} \label{anal}

The inner and outer ring spectra were extracted from the STIS 
spectral images. A portion of the G750M spectral image is shown in 
Figure~\ref{slit-pos}(b). The broad horizontal streak near the 
center is \Ha\ emission from the supernova debris 
($v_{\rm FWHM} \sim 2800 \kms$, \cite{wang96}; \cite{chug97}). 
The spectra of the WIR and EIR were obtained by intergrating $8-12$ rows 
of the image where the ring flux is recorded. 
The extraction height varied slightly with
the emission line species because of the different size and width
of the circumstellar ring at each emission line (Sonneborn et al. 1997; 
Oppenheimer 1999). 
The \Ha\ emission filling the length of the aperture
in Figure~\ref{slit-pos}(b) is from the diffuse LMC background
and was removed by linear interpolation above and below the extracted rows.
Diffuse LMC background emission is present in the Balmer lines and 
the [\oiii]\wll4959, 5007 transitions. 
It is a source of systematic error for the ring flux measurements, 
especially for the outer rings, as described below.
The integrated spectra were normalized to correct for the
height of the extraction slit, that is, the number of rows integrated.

The flux of each spectral feature was measured by fitting  
a Gaussian to the line profile. 
The background level for each line was fit by a quadratic 
over a 12 to 25 \AA\ ($40-50$ pixels) region of the spectrum centered on
the  feature being measured.
The best fit was determined by minimizing the total $\chi^{2}$ in which the 
Gaussian  and background coefficients were free parameters.
All the emission lines were fitted well by Gaussian profiles. 
We present the [\oiii]\wl5007 emission profile fitting as an example
in Figure~\ref{gaussfits}.
The reduced $\chi^2$, or the total $\chi^2$ divided by the number
of degrees of freedom, should be 1.0 for a perfect fit.
The reduced $\chi^{2}$ of our line fits were all within the range $0.7-2.0$.  
For each feature, the line flux and its error were computed from the 
best fit parameters and their associated uncertainties.
The statistical errors of all the line fluxes were adjusted so that the 
reduced $\chi^{2}$ for all fits is 1.0. 

As mentioned above, diffuse emission from the 30 Doradus region 
of the LMC filled the slit and contributed to 
the error of flux measurements of the rings. 
The diffuse background flux varied by a factor of $\sim 2$ along 
the slit in the immediate vicinity ($\pm$2\farcs5) of the supernova. 
The diffuse background level at a given position along the slit, 
determined by linear interpolation, could be measured only to 
an accuracy of $\sim$~15\%.
For the inner ring, the subtracted background was only a small fraction 
of the line emission ($<$5\% for \Ha\ and $<$10\% for [\oiii]\wl5007) 
and therefore the additional errors caused by 
background subtraction were negligible. 
On the other hand, the diffuse background levels were comparable to 
those of the measured outer ring fluxes for \Ha\ and [\oiii]\wll4959, 5007. 
The additional uncertainties due to the background subtraction 
were larger for the SOR than the NOR because the lower fluxes of the SOR. 
We included the extra uncertainty due to this background  
in the calculations of the total error in the outer ring flux measurements.

Results of the emission line flux measurements, normalized to \Ha, are 
tabulated in Table~\ref{tb-flux}. 
For the outer rings, the tabulated errors were determined by adding
the uncertainties in the background flux and 
the statistical uncertainties in quadrature. 
As discussed in \S\ref{spot3}, the EIR \Ha\ emission
overlaps with a weak redshifted high velocity emission originating from 
the interaction between the supernova debris and inner ring. 
The EIR \Ha\ flux listed in Table~\ref{tb-flux} has been corrected
for the small contribution from this ``hotspot.''
The tabulated fluxes are dereddened with 
E($B-V$) of 0.16 (\cite{fitz90}) and the extinction law of 
Cardelli, Clayton, \&~Mathis (1989) with an assumed R$_{V}$ of 3.1. 
The differences between the LMC extinction law and the Galactic 
law are negligible in the optical at low color excess and have been 
ignored in our analysis (\cite{fitz99}).

\section{RESULTS}

\subsection{Inner Ring}   \label{iring}

Forbidden line ratios were used to estimate the physical conditions in the 
circumstellar rings. 
We updated the methodology described by Osterbrock (1989) 
with multilevel model atoms and with more recent atomic data. 
We included five levels for \sii, whereas for \nii\ and \oiii\ we 
used six-level atoms. 
We included \oiii\ atomic data from Aggarwal (1993) 
and Galavis, Mendoza, \&~Zeippen (1997), 
the results of Stafford et al.\ (1994) and Galavis et al.\ (1997)
for \nii, and results from Keenan et al.\ (1993) and 
Ramsbottom, Bell, \&~Stafford (1996) for \sii. 
Previous studies established that gas with the range 
$n_{\rm e} \sim 10^3-10^4 \cm3$ are present in the inner ring 
(\cite{fran89}; \cite{lund96}; \cite{sonn97}). 
Within this density range, the [\nii] and [\oiii] line ratios,
\begin{displaymath}
R({\rm N~II}) = {(I(\lambda 6548)+I(\lambda 6583)) \over I(\lambda 5755) }
\ \  {\rm and}
\end{displaymath}
\begin{displaymath}
R({\rm O~III}) = {(I(\lambda 4959)+I(\lambda 5007)) \over I(\lambda 4363) },
\end{displaymath}
are not very sensitive to electron densities $n_{\rm e}$ and provide 
a good measure of the electron temperature $T_{\rm e}$. Similarly, the ratio 
\begin{displaymath}
R({\rm S~II}) = {I(\lambda 6716) \over I(\lambda 6731) }
\end{displaymath}
is a good diagnostic of $n_{\rm e}$ with a weak dependence 
on $T_{\rm e}$. 

The assumption inherent in the nebular analysis technique is
that the gas considered has a constant
$n_{\rm e}$ and $T_{\rm e}$ within the volume. 
This condition is not met in the \sn\ circumstellar gas
and therefore the physical conditions derived below in 
\S\ref{i-ratio} with the line ratios represent only an average range 
of $n_{\rm e}$ and $T_{\rm e}$ that can be present in the gas. 
The situation is further complicated by the uncertainty in 
the geometry of the circumstellar gas. 
LS00 showed that the same [\nii] and [\oiii] line fluxes observed
would imply a higher [\nii]- and [\oiii]-emitting gas density for the
case of a ionization-bounded ring than for
the density-bounded, or, ``truncated,'' case.
As a consistency check of the results obtained from the nebular analyses, 
we therefore constructed time-dependent photoionization/recombination 
models below in \S\ref{i-model} which incorporated the geometry effect 
and calculated $n_{\rm e}$ and $T_{\rm e}$ self-consistently.

\subsubsection{Temperature and Density from plasma diagnostics} \label{i-ratio}

In Figure~\ref{niiratio}, we plotted the relationship between 
the line ratio $R$(\nii) and $T_{\rm e}$ in our atomic model. 
Our calculated $R$(\nii) values were very similar to those computed 
by McKenna et al.\ (1996), as the only difference is that we 
used newer transition probabilities (\cite{gala97}).  
The electron densities considered in Figure~\ref{niiratio}, 
$n_{\rm e} = 1000$ and $4000 \cm3$, were the lower and upper density 
limits respectively, of the [\nii] emitting gas based on the modeling 
results of [\nii]\wl6583 \hst/WFPC2 inner ring images up to day~3478 at the 
EIR and WIR positions (LS00). 
The observed values of $R$(\nii) for the WIR and EIR regions, 
as well as the derived range on $T_{\rm e}$ as constrained by the 
density range under consideration, are also shown in Figure~\ref{niiratio}
and listed in Table~\ref{tb-nete}.

For the [\oiii] temperature, we calculated the relationship
between the line ratio $R$(\oiii) and $T_{\rm e}$ for electron
densities of 800 and $2400 \cm3$. 
The range of $n_{\rm e}$ considered for [\oiii] was slightly lower 
than that for [\nii], as suggested by an analysis of 
[\oiii]\wl5007 \hst/WFPC2 images (LS00).
$R$(\oiii) and $T_{\rm e}$(\oiii) are also listed in Table~\ref{tb-nete}. 
The uncertainty in the derived [\oiii] temperature was larger than
that for [\nii] because the [\oiii]\wl4363 emission was very weak. 
We were only able to derive a lower limit of the temperature for the EIR
region and we listed the 1-$\sigma$ lower limit in Table~\ref{tb-nete}.

We calculated the relationship between the line 
ratio $R$(\sii) and $n_{\rm e}$ for electron temperatures 5000 and 10,000~K. 
The temperature limits chosen were based on the results from an  
analysis of the day~2876 0\farcs45 diameter aperture \hst/FOS spectrum 
centered at p.a.=300\arcdeg\ of the inner ring (LS00).
The ring subsection sampled in the FOS spectrum lies close to the WIR 
region that is studied in this work. 
The derived range of the [\sii] density from $R$(\sii) is also listed 
in Table~\ref{tb-nete}.

The temperatures measured here for the WIR and EIR are close to those 
reported by LS00 for the inner ring at p.a.~$= 300\arcdeg$, 
which were $\sim 10,000$~K for [\nii], and $\sim 27,000$~K for [\oiii]. 
The slightly higher [\nii] temperature found at 
p.a.~$= 283\arcdeg$ (EIR) compared with p.a.~$= 103\arcdeg$ (WIR) 
appears to be real. 
This indicates that the [\nii]-emitting gas could 
be slightly denser at p.a.~$= 103\arcdeg$, 
which is consistent with the ongoing trend observed since day~3270 
from \hst/WFPC2 imaging (LS00). 
As discussed below, the density difference does not have to be large. 

\subsubsection{Consistency check from modeling} \label{i-model}

We constructed photoionization/recombination models for the inner
ring to check the results derived from the plasma diagnostic line 
ratios in the previous section. 
The code for the model and the input elemental abundances were the same as 
those used in Lundqvist (1999), except for the Si abundance which 
was reduced to Si/H$~=1\times10^{-5}$ (cf. LS00). 
Recently, the spectrum from the supernova breakout has 
been re-calculated with the multienergy group radiation hydrodynamics 
code {\it STELLA\/} and it is in good agreement with the
early (up to $\sim$~day 150) optical/UV light curves of \sn\ (\cite{blin00}).
That result will be used later in a more thorough investigation of the rings 
(P.\ Lundqvist, in preparation). 
Here we used the model ``{\it 500full1}'' burst spectrum of 
Ensman \&~Burrows (1992), which was accurate enough for our purposes.

Guided by the densities derived from the [\sii] line ratio, we computed 
models for atomic densities in the range $500 - 5000 \cm3$. 
The resulting electron density would be $\sim$ 17\% higher if the gas 
is fully ionized, but is probably the same as the atomic density 
in the epoch that we are considering. 
The inner radius of the ring is assumed to be $6.3\EE{17}$~cm, as in
Lundqvist \&~Fransson (1996). 
We modeled the temperature sensitive ratios $R$(\nii) and $R$(\oiii) 
as a function of atomic density (Figure~\ref{iring-mod}). 
Although the light travel time differences are not as 
important across the inner ring as for the outer rings (see below),
we included this effect in our models. 
The line ratios were calculated for two cases: 
an ionization-bounded model, and a model where we simulated a 
density-bounded situation by including only the innermost 25\% of 
the ionized gas, which we referred to as ``25\% truncation''.
$R$(\nii) quickly increases from low values ($\sim 15-25$) 
to very large values ($\ga 100$) 
at a density in the range $1000 - 3000 \cm3$, the exact value of which 
depends on the degree of truncation.  
For example, $R$(\nii) = 50 is reached at 
$\sim 1700 \cm3$ for the $25\%$ truncation model, and at $\sim 2700 \cm3$ 
for the ionization-bounded case. 

In our model, the observed $R$(\nii) ratios in the inner ring 
are actually very sensitive to density.
This is contrary to what is expected in the standard nebular 
analysis where $R$(\nii) is used to determine electron 
temperature because it is relatively insensitive to variations 
in electron densities. 
The reason for this discrepency is that the temperature of 
the emitting gas at a specific epoch in the \sn\ inner ring depends 
on the density of the gas, which drops faster for the higher density
component. This translates directly into different line ratios
for different density components within the gas. 
Only models with an average temperature of the line emitting gas 
similar to the temperature given by the nebular analysis will 
give the correct line ratio.

On the other hand, the degree of truncation has only minor effects on 
the dependence of the [\oiii] line ratio with density. 
Assuming these two cases represent the range of 
possible models (LS00), we derived the [\nii] density of the WIR 
region to be $2400 - 3500 \cm3$ and the [\oiii] density to be 
$1400 - 2100 \cm3$. For the EIR region,  
the corresponding numbers were $1700 - 3000 \cm3$ and $\la 2100 \cm3$,
respectively. 
This suggests that the [\sii], [\nii], and [\oiii] emission come 
from regions of progressively lower density, consistent with
the findings of LS00. 
Moreover, the densities derived in this paper (day~4282) are 
somewhat lower than those in LS00 (day~3000).
This is consistent with the trend that gas components of progressively lower 
density dominate the emission for a particular transition 
at later and later times. 
The higher the ionization potential of a line, the lower the density of the
gas that emits it at a specific epoch. This 
has the interesting effect that the density components dominating 
the [\oiii] emission in the epoch studied in LS00 ($\sim$~day 3000), 
could be the same as those dominating the [\nii] emission in 
this analysis (day~4282).

\subsection{Outer Rings} \label{oring}

With the emission from the outer rings much weaker than that
from the inner ring, the uncertainties in the outer ring fluxes are 
much larger than those for the inner ring (cf. Table~\ref{tb-flux}). 
Both [\nii]\wl5755 and [\oiii]\wl4363 are absent, or very close 
to the noise limit, in our data for the outer rings. 
Therefore we cannot obtain a meaningful estimate of the 
electron temperature directly from the line ratios $R$(\nii) and $R$(\oiii).
The [\sii]\wll6716, 6731 lines, on the other hand, are both relatively 
strong and do not suffer from systematic errors in background 
subtraction (cf.\ \S\ref{anal}).
Using the $R$(\sii) method (\S\ref{iring}), 
we arrived at a 1-$\sigma$ upper limit on the electron density 
of the NOR (SOR) of $n_{\rm e} \la 2200~(1600) \cm3$, 
allowing for a $T_{\rm e}$ between 5000~K and 10000~K. 
The derived outer ring densities are about a factor of two lower 
than the inner ring. 
The slightly higher electron density for the NOR, which 
lies further away from us than the SOR, may indicate that the effect
of light travel time is important for the outer rings.

As a consistency check, we also computed photoionization/recombination 
models for the outer rings to model the observed [\sii]/\Ha, [\nii]/\Ha, and
[\oiii]/\Ha\ flux ratios. 
The code and elemental abundances were the same as those 
used for the inner ring analysis (\S\ref{i-model}). 
We adopted $2\EE{18}$~cm as the distance of the outer rings 
from the supernova. 
This means that there is a $\sim 900$ day difference in age of
the emission from the observed positions on the SOR and NOR (\S\ref{obs}). 
We took this different light travel time into account in our models. 
Guided by the densities derived from $R$(\sii) for the outer rings, 
we ran models for atomic densities in the range $500 - 2000\cm3$. 

As discussed in Lundqvist \& Fransson (1996), the degree of ionization
of the circumstellar gas after shock breakout is sensitive to the distance 
of the gas to the supernova. 
It is particularly important for the ``{\it 500full1}'' model
by Ensman \&~Burrows (1992) because of the smaller number of high
energy photons in the burst spectrum. 
At a distance 3.2 times farther away from the supernova than 
the inner ring, the inner part of the outer rings was 
dominated by \civ, \niv, and \oiv\ shortly after breakout, 
instead of the higher ionized species such as \ion{N}{6}, and \ovii\ 
that were present in the inner ring, at the density range that we
are considering (\cite{lund96}). 
We calculated that the maximum temperature reached in the outer rings 
from the initial UV flash was only $\sim 7.8\EE4$ K, 
a factor of $\sim 2$ smaller than that in the inner ring. 
At $\sim$~day 3500, roughly the mean epoch of emission for the SOR and 
NOR positions we observed, for an atomic density of $1400 \cm3$,
the maximum temperature in the outer rings would be $\sim 1.8\EE4$~K, 
and the dominant ions \cii, \nii, and \oii. 
For a density of $500 \cm3$, the dominant ions would be  
\ciii, \niii, and \oiii\ instead, and the temperature would be in the range 
$(1.0-4.5)\EE4$~K. 
In Figure~\ref{oring-mod} we show the modeled emissions in 
[\sii]\wll6716, 6731, [\nii]\wll6548, 6583, and [\oiii]\wll4959, 5007, 
normalized to \Ha\ as functions of atomic density. 
Similar to the inner ring, we considered two limiting cases
for the outer ring modeling: an ionization-bounded 
model and a 25\% truncation model (cf.\ \S\ref{i-model}).
We plotted the observed normalized fluxes [\nii]/\Ha, [\oiii]/\Ha, 
and [\sii]/\Ha\ for both the SOR and NOR regions for comparison with 
the models. 

For the NOR, the density implied by the [\sii]/\Ha\ flux ratio
in the model ($1000-2000 \cm3$) was indeed consistent with that 
indicated by the line ratio $R$(\sii) in Table~\ref{tb-nete}. 
We concluded that the observed NOR [\sii] lines 
could be interpreted with a model where the input S abundance 
was the same as that of the inner ring, i.e., S/H~$\sim 6\EE{-6}$. 
Similarly, Figure~\ref{oring-mod} also suggested that the NOR 
[\nii] and [\oiii] emissions were consistent with a model where 
the emitting gases ($n_{\rm e} \sim 1300 \cm3$) were less dense than 
those in the inner ring, but with the same N and O abundances. 
Our results favor a coeval scenario for the inner and outer rings, 
as argued for in Crotts \& Heathcote (2000). 
Due to large uncertainties in the measured fluxes, caused largely
by uncertainties in the background subtraction, we were 
unable to determine whether the NOR is ionization- or density-bounded.  

For the SOR, the [\sii] density estimate from our model
($700-1500 \cm3$) was again consistent with that from the
$R$(\sii) calculations. 
However, even for the 25\% truncation model, the modeled [\nii] 
emission from the SOR is $\ga 10\%$ higher than the observed value. 
In fact, only truncated models (and possibly with even more truncation,
such as a $10\%$ truncation model) can possibly fit the results, if we 
assume that the N abundance in the SOR is similar to that 
in the inner ring. 
The SOR [\nii] emission then implies a density $\la 1000 \cm3$.
However, the [\oiii] density inferred for the SOR in 
Figure~\ref{oring-mod} is $\sim 1500 \cm3$.
This result, that the [\oiii]-emitting region has higher density
than that of the [\nii]-emitting region in the SOR, 
is opposite to what is expected (cf.\ \S\ref{i-model}). 
We note though, that the modeled ratio of [\nii]/[\oiii]
is the same as the observed SOR value for a density of $\sim 1000 \cm3$, 
which could either indicate a lower overall metal abundance in the SOR 
than in the NOR (although with similar relative abundances of N and O), 
or that the adopted value of the LMC \Ha\ background in the SOR
which we subtracted was too low. 
As described in \S\ref{anal}, the SOR region is more vulnerable to 
background subtraction errors than the NOR because it is fainter.
The observed SOR [\nii]/\Ha\ flux ratio could be brought to agreement
with the model if the real LMC \Ha\ background at the position of
SOR was $\sim$~10\% larger than the value we have adopted. 
Despite this problem, we note that the [\oiii]/H$\alpha$ ratio is very 
sensitive to the density, placing a lower limit on the density of 
the emitting regions in both of the outer rings at $\sim 1000 \cm3$.

The densities we derive, $\sim 1000-2000 \cm3$, are somewhat larger than 
estimated by Panagia et al.\ (1996), who found $800 \cm3$ on day 2887 
for a single position of the northern outer ring. 
This is not necessarily a real contradiction, since the UV lines 
studied by Panagia et al.\ probe mainly hot gas ([\niii], \ciii], [\oiii]), 
which, in order to stay hot at these late epochs cannot be dense. 
However, we do not feel that their result, that the N abundance in 
the outer rings is a factor of $\sim 2-3$ smaller than the inner ring,
is conclusive.  
A time dependent, and self-consistent
analysis is required to study the extreme non-equilibrium conditions 
encountered in the rings at late epochs. 

We close this discussion by noting that creating a density 
of $\sim 1000-2000 \cm3$ at the position of the outer rings poses a real 
challenge to models of the formation of the ring system. 
For example, it is a factor $\sim 10$ higher than indicated by the models 
of Blondin \&~Lundqvist (1993) and Martin \&~Arnett (1995). 
Multidimensional models including photoionization by the progenitor 
have not yet demonstrated an improvement of this situation 
(e.g., \cite{chev00}).
High quality spectra with good S/N are needed to accurately 
determine the temperature, ionization and abundances of the gas 
in the outer rings. 

\subsection{Spot 3 discussion}  \label{spot3}

\begin{figure}
\plotone{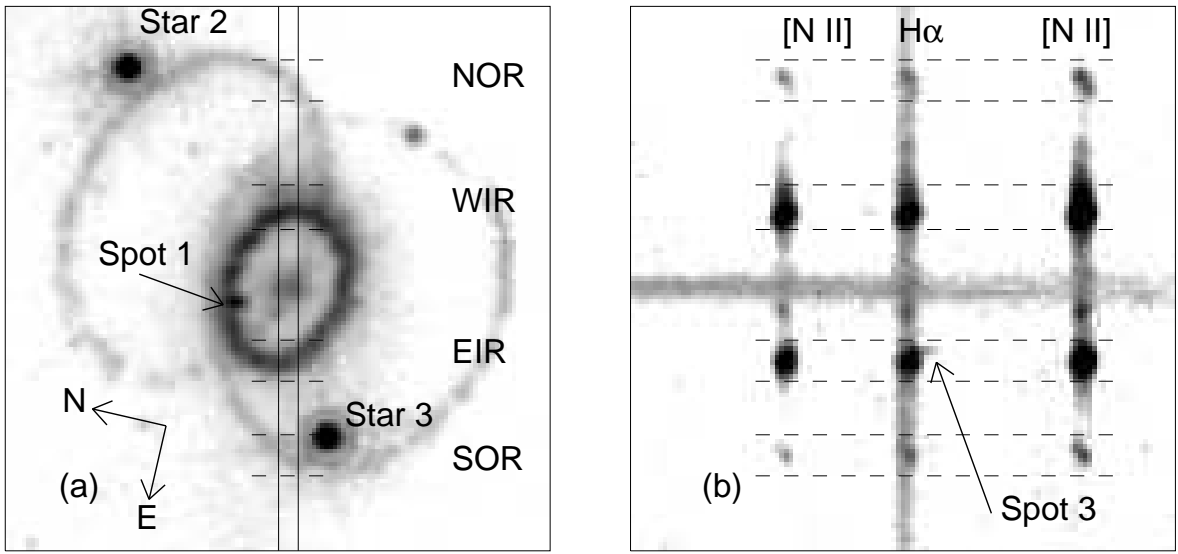}
\caption{(a) The size and orientation of the STIS aperture are 
shown on a 1999 January WFPC2 image of \sn. The image size is 
5\farcs8 square. The regions of the west inner ring (WIR),
east inner ring (EIR), north outer ring (NOR), and south outer
ring (SOR) spectra extractions are also shown by horizontal dashed lines; 
(b) Section of the STIS 
G750M (6581) spectral image showing the emission from \Ha\ and 
[\nii]\wll6548, 6583. The spatial scale is identical to (a).  The dashed 
lines show the spectral extraction regions for the inner and outer ring 
spectra. The arrow points to the redshifted \Ha\ emission that indicates 
the presence of shock interaction. Its position also indicates that this 
emission originates on the inner edge of the ring.}
\label{slit-pos}
\end{figure}

\begin{figure}
\plotfiddle{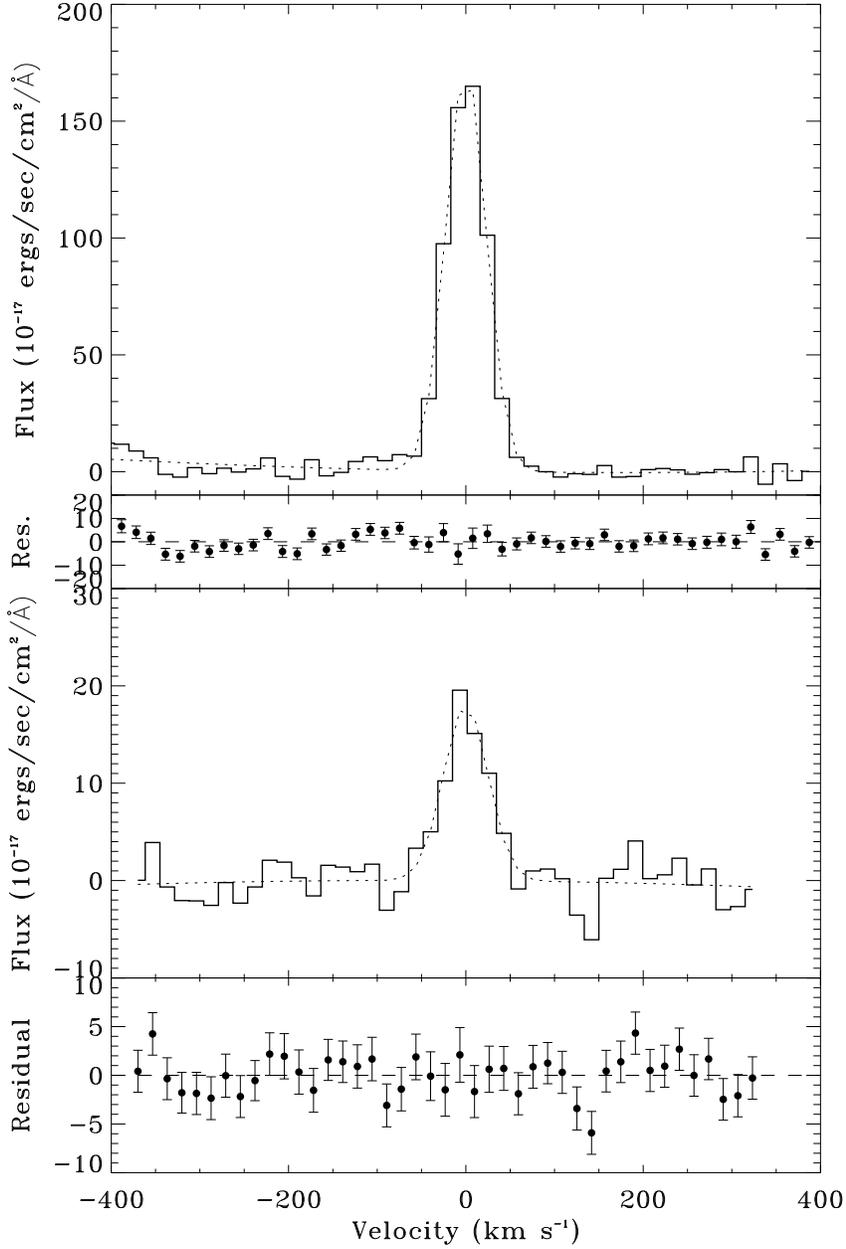}{16cm}{0}{90}{90}{-180}{-20}
\caption{Sample gaussian profile fitting for the [\oiii] \wl5007 emission.
The top panel shows the [\oiii] \wl5007 emission from the EIR region 
(thick line) along with the best $\chi^2$ fit 
(broken line). The reduced $\chi^2$ of the fit is 1.9. 
The bottom panel shows the [\oiii] \wl5007 emission from the SOR 
region (thick line) along with the best $\chi^2$ fit 
(broken line). The reduced $\chi^2$ of the fit is 1.0.}
\label{gaussfits}
\end{figure}

\begin{figure}
\plotone{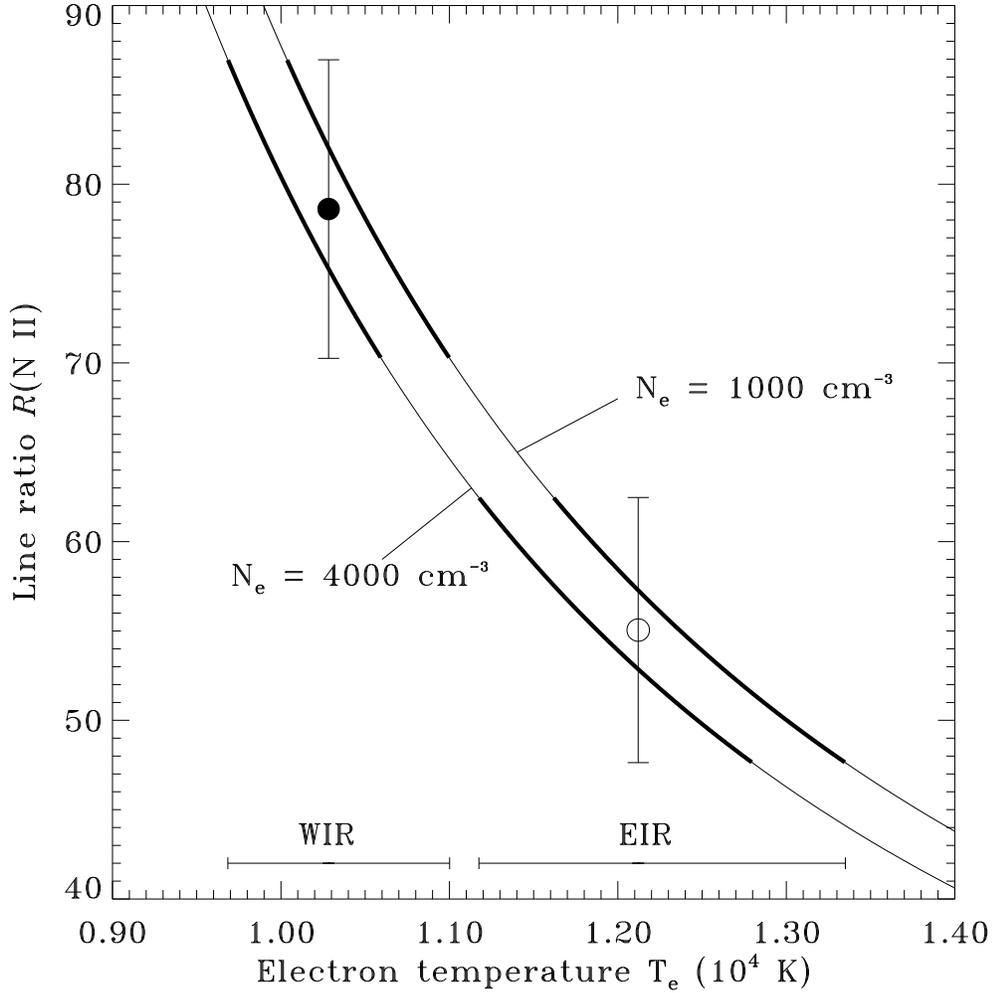}
\caption{The theoretical [\nii] line ratio (\wl6548 + \wl6583)/\wl5755 as a 
function of electron temperature is shown for two electron densities 
(1000 and 4000$\cm3$). 
The observed [\nii] line ratios for the WIR (filled circle) and 
EIR (open circle) inner ring areas and the implied electron 
temperatures for the two regions are also shown.
}
\label{niiratio}
\end{figure}

\begin{figure}
\plotone{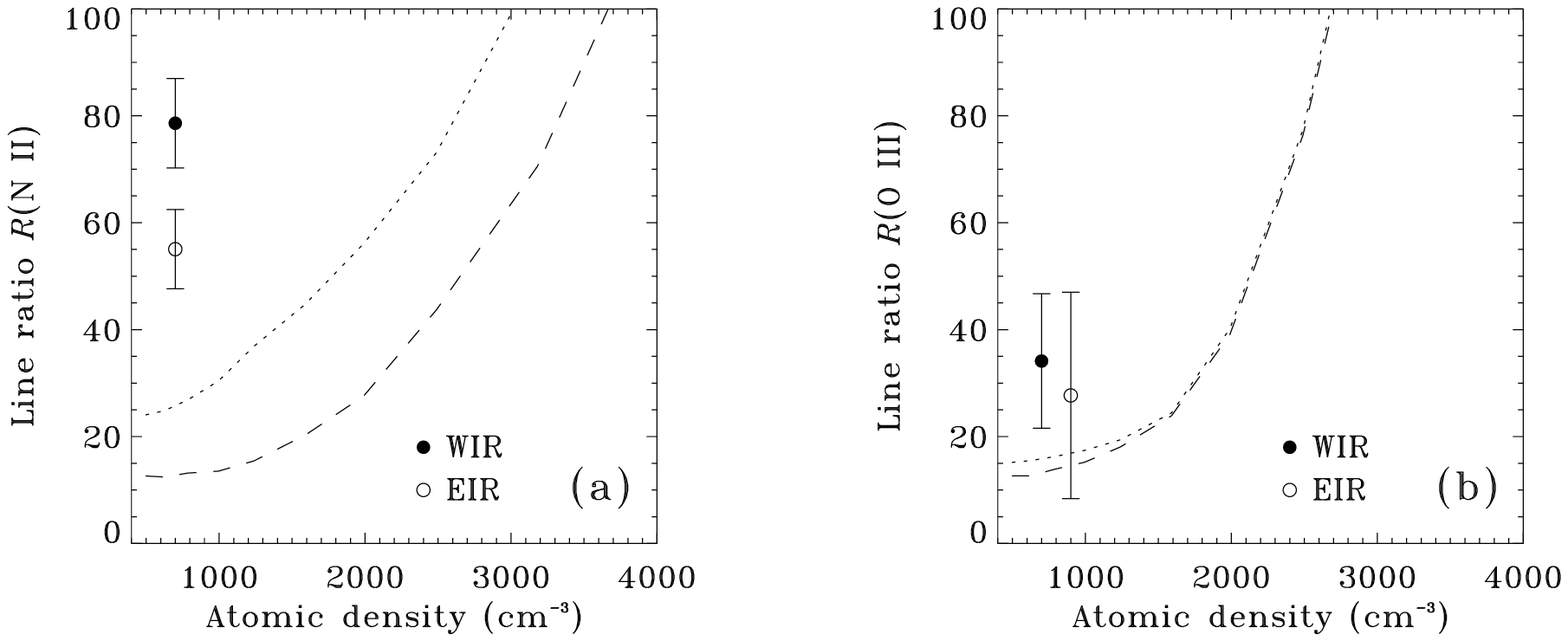}
\caption{(a) The computed model [\nii] line ratio (\wl6548 + \wl6583)/\wl5755 
is plotted against the atomic density for the \sn\ inner ring gas with
an ionization bounded model (dash) and a 25\% truncation density
bounded model (dotted); (b) Similar as (a), for the model
[\oiii] line ratio (\wl4959 + \wl5007)/\wl4363. 
The measured [\nii] and [\oiii] line ratios from the west (WIR, filled circle)
and east (EIR, empty circle) inner ring areas are also shown.}
\label{iring-mod}
\end{figure}

\begin{figure}
\plotone{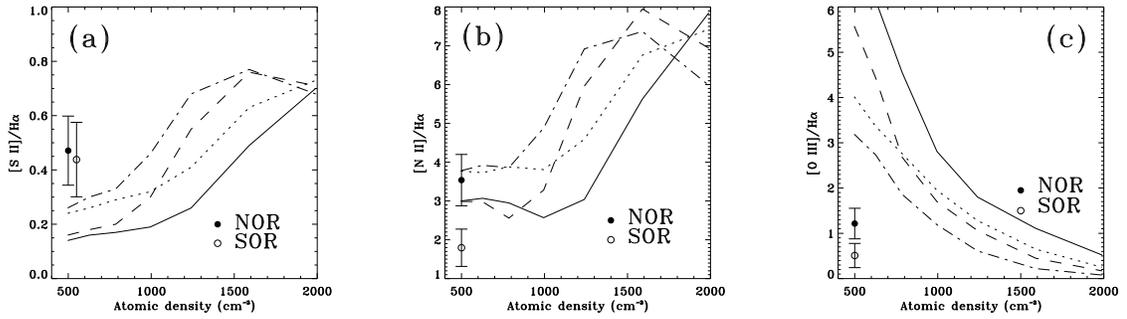}
\caption{(a) The computed model [\sii] line flux (\wl6716 + \wl6731)/\Ha\ is 
plotted against the atomic density for the NOR gas 
with an ionization bounded model (dotted) and a 25\% truncation density
bounded model (solid), along with the model line fluxes for the 
SOR gas with the ionization bounded model (dot-dash) 
and the 25\% truncation model (dash); 
(b) Similar as (a), for the model [\nii] line flux
(\wl6548 + \wl6583)/\Ha; 
(c) Similar as (a), for the model [\oiii] line flux
(\wl4959 + \wl5007)/\Ha; 
The measured line fluxes from the 
northern (NOR, filled circle) and southern (SOR, empty circle) 
outer ring areas are also shown.}
\label{oring-mod}
\end{figure}

The first evidence for the collision between the supernova ejecta
and the inner circumstellar ring came in 1997 April from 
the STIS observations of a \Ha\ emission feature blueshifted up to
$\sim250\kms$ that was projected near the inside edge of the spectral
image of the ring (\cite{sonn98}). 
Subsequent photometric observations with \hst/WFPC2 (\cite{garn00a}) and 
spectroscopic observations with STIS (\cite{mich00}) of this hotspot
(Spot~1) demonstrate continued brightening plus emissions 
in other species such as [\oi], [\nii], and [\oiii]. 
The appearance of Spot~1 was explained as the result of the 
collision of the outer shock from the \sn\ debris with a dense
knot of gas protruded inward towards the supernova from the ring. 
Additional spots were predicted to appear once 
the debris collides with the rest of the ring. 
The presence of a new source of radiation was first reported in 
the CTIO \hei\ 1.08$\mu$m image in 1999 December 25 by 
Bouchet et al.\ (2000).
Subsequent STIS (Maran, Pun, \&~Sonneborn 2000) and WFPC2 
(Garnavich, Kirshner, \&~Challis 2000b) imaging showed that 5 additional 
bright spots have developed around the inner ring. 
In Figure~\ref{slit-pos}(b), we point out a redshifted 
component in \Ha\ in the G750M (6581) spectral image
which was first identified in this data by Lawrence \&~Crotts (2000).
This high-velocity \Ha\ emission originates from the inner edge of the
EIR region. 
No similar Doppler-shifted emission is observed in our data
for any other species.
The position of this redshifted \Ha\ flux coincides with that of Spot 3
(or HS2-106, using the nomenclature of \cite{lawr00}), 
one of the recently discovered new regions of brightening on the
\sn\ inner ring reported by Garnavich et al.\ (2000b).
Measurements of Spot 3 in the 2000 February 2 (day~4727) WFPC2 images 
gave a position of 0\farcs70 away from the center of the \sn\ debris
and a position angle of 106\arcdeg. 
This is the start of the full scale collision between the supernova ejecta
and the inner circumstellar ring, as also pointed out by 
Lawrence et al.\ (2000).

\begin{figure}
\plotone{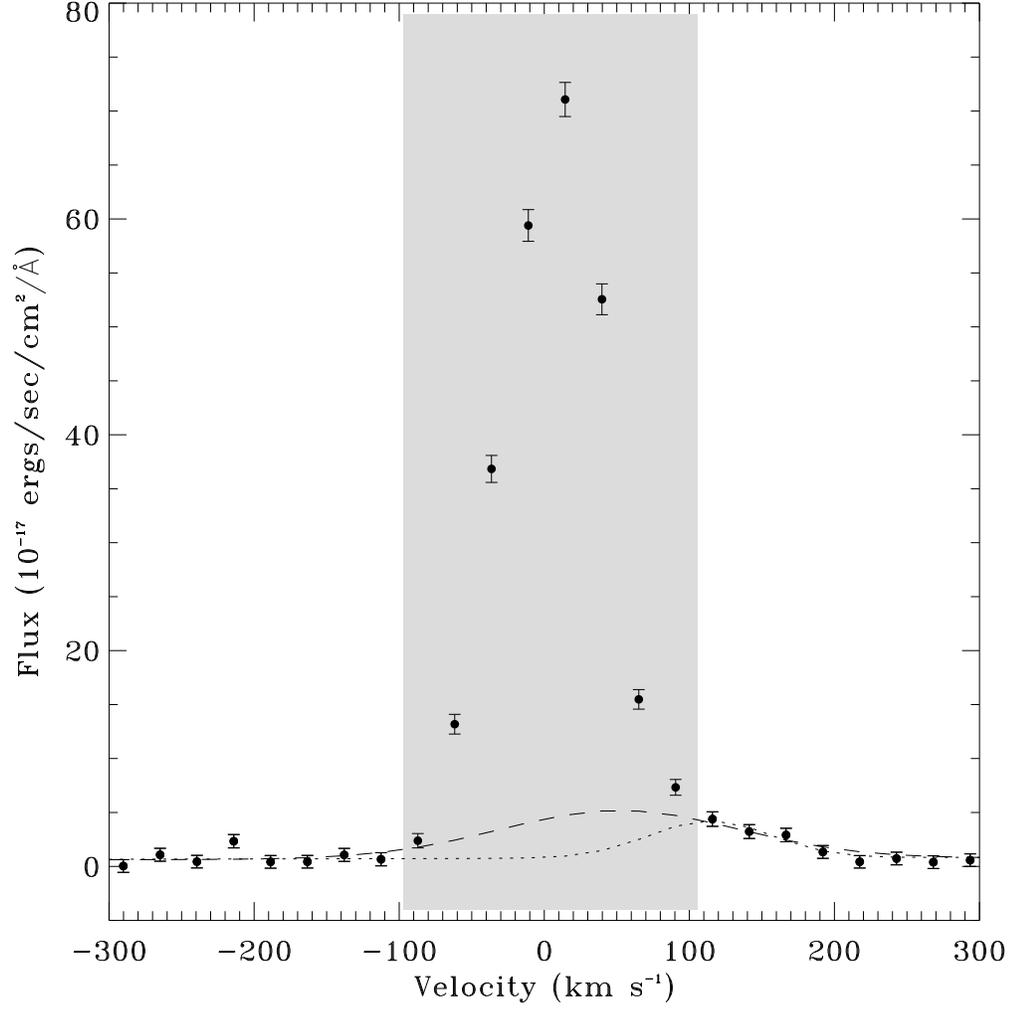}
\caption{The \Ha\ profile from the east section of the inner ring. 
The shaded area represents the region where the radiation is 
dominated by the ring and not used for the fitting. 
The redshifted emission from Spot 3 is fitted 
by a Gaussian for both the cases where the width of the emission is 
a free parameter (dot) and where the FWHM of the emission is set 
to be 195$\kms$ (dash).}
\label{ha-spot}
\end{figure}

The combined \Ha\ spectrum of the Spot 3 and the overlapping portion of 
the EIR is extracted in a similar manner as that described in \S2, 
except that fewer rows ($n=4$) are integrated for 
the output spectrum.
The extracted \Ha\ profile for Spot 3 and the ring is
shown in Figure~\ref{ha-spot}.
Redshifted emission from Spot 3 is clearly observed up to 
$\sim +175 \kms$ while the blueshifted component cannot be
identified because the extracted spectrum is dominated by 
the EIR flux that fills the width of the aperture between $\pm100\kms$.
We fitted the rest of the \Ha\ emission with a Gaussian function 
and a linear background. 
The best-fit Gaussian is shown by the dotted line in 
Figure~\ref{ha-spot}. The \Ha\ flux from Spot 3 derived from the 
best-fit parameters is $(8\pm4)\EE{-17} \ergcms$ (no reddening 
correction).
With only the redshifted profile available for fitting, the 
FWHM of the emission is poorly constrained to be $100 \pm 50 \kms$. 
On the other hand, the emission-line width of Spot~1 in \Ha\ measured 
from much better S/N STIS observations gives a FWHM width of $253 \pm 4 \kms$
(\cite{mich00}). Subsequent monitoring of the Spot~1 \Ha\ radiation
profile suggests that the width of the radiation remains constant
with time (C.S.J.\ Pun, in preparation). 
If we assume that the intrinsic width of Spots 1 and 3 are identical, 
then we expect the {\it observed\/} FWHM of the Spot 3 profile 
to be $195\kms$, after correcting for the
different position angles of the two spots on the ring. 
With the profile width fixed, we determined a more accurate
Spot 3 flux to be $(2.0\pm0.6)\EE{-16} \ergcms$ 
(no reddening correction). 
The best fit profile is shown with dashed line in Figure~\ref{ha-spot}. 
The portion of this radiation ($\sim$68\%, or $1.4\EE{-16} \ergcms$) 
that lies within the 0\farcs2 aperture has been removed for the 
determination of the EIR \Ha\ inner ring flux (cf.\ \S\ref{anal}).
The Spot~3 flux measured in this paper was consistent with
the value previously reported using the same data by Lawrence et al.\ (2000).


Using the CALCPHOT task in the SYNPHOT package of the STSDAS
data reduction software, we converted the Spot 3 flux in
1998 November to an equivalent STMAG magnitude of 21.7 (0.4). 
If we assume the rate of brightness increase for Spot~3 is 
the same as that of Spot~1, $2.09 \pm 0.11$~mmag~day$^{-1}$
(\cite{garn00a}), then the extrapolated brightness for Spot~3 
on day~4727 will be 20.7 (0.4). 
Our predicted flux was $\sim 1$~mag fainter than the measured 
WFPC2 \Ha\ photometric result of 19.6 ($1.6\EE{-15} \ergcms$)
on day~4727 (\cite{garn00b}), probably caused by the fact that 
Spot~3 is not centered within the STIS aperture in our data. 

Finally, we comment on the monitoring
of the interaction between the \sn\ debris and the circumstellar ring. 
While difference imaging in \Ha, [\nii], and \hei\ 1.083$\mu$m, where
the newly-formed spots have the greatest contrast above the 
equatorial inner ring, may be a sensitive technique (\cite{lawr00}), 
it is susceptable to problems in separating the flux from the newly
formed spots from the noisy background. 
It is not surprising that by separating the emission of 
the slow-moving ring ($v\sim10\kms$) from that of the 
interacting spots ($v\sim250\kms$), radiation from Spot 3 could be 
detected in the STIS data at least one year before discovery by
imaging, and over six months before it could be identified in 
pre-discovery images. 
This indicates that high resolution spectroscopic 
monitoring provides the best technique for discovering new areas
of shock interaction in \sn.

\acknowledgements
We are grateful to Peter Garnavich and Arlin Crotts for useful discussions. 
We thank Don Lindler, Phil Plait, and the Goddard STIS team for 
their help with data reduction, and the Supernova Intensive Study 
(SINS, PI: Robert Kirshner) team for access to their unpublished data.  
This work was supported by NASA Guaranteed Time Observer funding to the STIS 
Science Team (PI: Bruce Woodgate).
C.S.J.P.\ acknowledges funding by the STIS IDT through the National
Optical Astronomy Observatories. 
P.L.\ acknowledges support from the Swedish Natural Science Research Council 
and the Swedish National Space Board.

\clearpage

\begin{deluxetable}{cccl}
\tablewidth{0pc}
\tablecaption{
1998 November 14 (day 4282) STIS Observations
}

\tablehead{
\colhead{} & 
\colhead{Wavelength} &
\colhead{Exposure time} &
\colhead{} \\
\colhead{Gratings} & 
\colhead{Range (\AA)} &
\colhead{(sec)} &
\colhead{Emission lines observed} \\
}
\startdata
G750M (6581) & 6295~-- 6867 & 4750 & [\oi] 6300,64, \Ha, [\nii] 6548,83, \nl 
             &              &      & \hei \ 6678, [\sii] 6716,31 \nl
G750M (5734) & 5448~-- 6020 & 5250 & [\nii] 5755, \hei \ 5876 \nl
G430M (4961) & 4818~-- 5104 & 5750 & \Hb, [\oiii] 4959, 5007 \nl
G430M (4451) & 4308~-- 4594 & 7000 & \Hg, [\oiii] 4363 \nl
G430M (3423) & 3280~-- 3566 & 7000 & \nodata \nl
\enddata
\label{tb-obs}
\end{deluxetable}

\clearpage
\begin{deluxetable}{clllllc}

\tablewidth{0pc}

\tablecaption{Normalized and dereddened emission line fluxes from the 
inner and outer rings}
\tablehead{
\colhead{}    & \colhead{}                & \colhead{} & \colhead{} & 
\colhead{} & \colhead{} & \colhead{Extinction} \\
\colhead{Ion} & \colhead{$\lambda$ (\AA)} & \colhead{WIR\tablenotemark{a}} & 
\colhead{EIR\tablenotemark{b}} &
\colhead{NOR\tablenotemark{c}} &
\colhead{SOR\tablenotemark{d}} &
\colhead{Correction\tablenotemark{e}}
}
\startdata
\Hg       & 4340.46 & 0.155 $\pm$ 0.011 & 0.172 $\pm$ 0.014 & 0.333 $\pm$ 0.176 & 0.186 $\pm$ 0.103 & 1.850 \nl
[\oiii]   & 4363.21 & 0.015 $\pm$ 0.006 & 0.021 $\pm$ 0.015 & \nodata           & \nodata           & 1.843 \nl
\Hb       & 4861.32 & 0.352 $\pm$ 0.013 & 0.319 $\pm$ 0.019 & 0.514 $\pm$ 0.175 & 0.460 $\pm$ 0.206 & 1.702 \nl
[\oiii]   & 4958.91 & 0.136 $\pm$ 0.011 & 0.140 $\pm$ 0.016 & 0.284 $\pm$ 0.113 & 0.165 $\pm$ 0.110 & 1.679 \nl
[\oiii]   & 5006.84 & 0.384 $\pm$ 0.015 & 0.444 $\pm$ 0.024 & 0.932 $\pm$ 0.290 & 0.344 $\pm$ 0.227 & 1.668 \nl
[\nii]    & 5754.59 & 0.060 $\pm$ 0.006 & 0.085 $\pm$ 0.012 & 0.198 $\pm$ 0.096 & \nodata           & 1.543 \nl
\hei      & 5875.65 & 0.054 $\pm$ 0.004 & 0.056 $\pm$ 0.008 & \nodata           & \nodata           & 1.528 \nl
[\oi]     & 6300.30 & 0.154 $\pm$ 0.007 & 0.144 $\pm$ 0.012 & \nodata           & \nodata           & 1.480 \nl
[\oi]     & 6363.78 & 0.061 $\pm$ 0.006 & 0.044 $\pm$ 0.006 & \nodata           & \nodata           & 1.474 \nl
[\nii]    & 6548.05 & 1.189 $\pm$ 0.024 & 1.202 $\pm$ 0.045 & 1.062 $\pm$ 0.216 & 0.395 $\pm$ 0.122 & 1.454 \nl
\Ha       & 6562.82 & 1.000             & 1.000             & 1.000             & 1.000             & 1.453 \nl
[\nii]    & 6583.45 & 3.534 $\pm$ 0.084 & 3.486 $\pm$ 0.143 & 2.476 $\pm$ 0.470 & 1.399 $\pm$ 0.378 & 1.451 \nl
\hei      & 6678.15 & 0.014 $\pm$ 0.004 & 0.012 $\pm$ 0.005 & \nodata           & \nodata           & 1.441 \nl
[\sii]    & 6716.44 & 0.208 $\pm$ 0.006 & 0.201 $\pm$ 0.012 & 0.261 $\pm$ 0.073 & 0.243 $\pm$ 0.086 & 1.437 \nl
[\sii]    & 6730.82 & 0.347 $\pm$ 0.009 & 0.338 $\pm$ 0.018 & 0.210 $\pm$ 0.086 & 0.196 $\pm$ 0.072 & 1.436 \nl 
\enddata
\tablenotetext{a}{Dereddened WIR H$\alpha = (71.56\pm 0.95) \times 10^{-16} \ergcms$}
\tablenotetext{b}{Dereddened EIR H$\alpha = (41.12\pm 1.16) \times 10^{-16} \ergcms$}
\tablenotetext{c}{Dereddened NOR H$\alpha = (3.59\pm 0.51) \times 10^{-16} \ergcms$}
\tablenotetext{d}{Dereddened SOR H$\alpha = (4.01\pm 0.64) \times 10^{-16} \ergcms$}
\tablenotetext{e}{E($B-V$) = 0.16, and the extinction correction law
of Cardelli et al.\ (1989) and R$_{V}$ = 3.1.}
\label{tb-flux}
\end{deluxetable}
\clearpage

\begin{deluxetable}{lcccc}
\tablewidth{0pc}

\tablecaption{Plasma Diagnostics}
\tablehead{
\colhead{Parameter} & \colhead{WIR} & \colhead{EIR} & \colhead{NOR} &
\colhead{SOR}  \nl
}
\startdata
$R$(\nii)                     & $78.6 \pm 8.4$         & $55.0 \pm 7.4$        & \nodata         & \nodata \nl
$T_{\rm e}$(\nii) (K)         & $10300^{+700}_{-600}$  & $12100^{+1200}_{-900}$& \nodata         & \nodata \nl
$R$(\oiii)                    & $34.1 \pm 12.6$        & $27.7 \pm 19.3$       & \nodata         & \nodata \nl
$T_{\rm e}$(\oiii) (K)        & $21400^{+9400}_{-3700}$& $>$17700              & \nodata         & \nodata \nl
$R$(\sii)                     & $0.599 \pm 0.020$      & $0.596 \pm 0.041$     & $1.16 \pm 0.45$ & $1.20 \pm 0.42$ \nl
$n_{\rm e}$(\sii) (cm$^{-3}$) & $3880^{+1150}_{-1060}$ & $4000^{+2410}_{-1510}$& $<$2200         & $<$1600         \nl
\enddata
\label{tb-nete}
\end{deluxetable}

\clearpage

\clearpage

\clearpage

\clearpage

\clearpage

\clearpage

\end{document}